\documentclass[a4paper, 10pt, conference]{ieeeconf}      

\IEEEoverridecommandlockouts                              

\usepackage{comment}
\usepackage{hyperref}
\usepackage{graphicx}
\usepackage{color}
\usepackage{amsmath}
\usepackage{amssymb}
\usepackage{xcolor}
\usepackage{color}
\usepackage{float}
\usepackage{subcaption}
\usepackage{listings}
\usepackage{booktabs}
\usepackage[T1]{fontenc}

\colorlet{punct}{red!60!black}
\definecolor{background}{HTML}{EEEEEE}
\definecolor{delim}{RGB}{20,105,176}
\colorlet{numb}{magenta!60!black}

\lstdefinelanguage{json}{
    basicstyle=\normalfont\ttfamily,
    numbers=left,
    numberstyle=\scriptsize,
    stepnumber=1,
    numbersep=8pt,
    showstringspaces=false,
    breaklines=true,
    frame=lines,
    backgroundcolor=\color{background},
    literate=
     *{0}{{{\color{numb}0}}}{1}
      {1}{{{\color{numb}1}}}{1}
      {2}{{{\color{numb}2}}}{1}
      {3}{{{\color{numb}3}}}{1}
      {4}{{{\color{numb}4}}}{1}
      {5}{{{\color{numb}5}}}{1}
      {6}{{{\color{numb}6}}}{1}
      {7}{{{\color{numb}7}}}{1}
      {8}{{{\color{numb}8}}}{1}
      {9}{{{\color{numb}9}}}{1}
      {:}{{{\color{punct}{:}}}}{1}
      {,}{{{\color{punct}{,}}}}{1}
      {\{}{{{\color{delim}{\{}}}}{1}
      {\}}{{{\color{delim}{\}}}}}{1}
      {[}{{{\color{delim}{[}}}}{1}
      {]}{{{\color{delim}{]}}}}{1},
}

\title{\LARGE \bf A Cyber-Physical Machine Tool Framework with a Real-Time Machining Process Digital Twin}

\author{Khalil Chakal$^{1}$, Tero Kaarlela$^{1}$, Jose Outeiro$^{2}$ and Carlos Andrade$^{2}$
\thanks{$^{1}$Materials and Mechanical Engineering, Faculty of Technology, University of Oulu, Oulu, Finland.
        {\tt\small khalil.chakal@oulu.fi}}%
\thanks{$^{2}$Digital Engineering for Advanced Manufacturing Laboratory (DEAM Lab), Center for Precision Metrology, Department of Mechanical Engineering and Engineering Science, University of North Carolina at Charlotte, 9201 University City Blvd., Charlotte 28223, NC, USA}
\thanks{This work has been submitted to the IEEE for possible publication.
Copyright may be transferred without notice, after which this version may
no longer be accessible.}}

\begin{document}

\maketitle
\thispagestyle{empty}
\pagestyle{empty}

\begin{abstract}

Digital Twins (DTs) have emerged as a key technology for improving the monitoring, optimization, and automation of manufacturing systems. However, existing Cyber-Physical Machine Tool (CPMT) implementations primarily represent the machine tool, while the machining process remains only partially synchronized with its physical counterpart. This paper extends a previously presented CPMT framework by introducing a hierarchical DT framework that simultaneously maintains DTs of both the machine tool and the machining process. The proposed framework integrates real-time CNC operational data, a voxel-based workpiece representation, synchronized process vibration measurements, and a persistent part DT repository for process replay, traceability, and future synthetic data generation. 

Experimental evaluation demonstrated real-time operation at a 20 Hz machining-state update rate, interactive visualization exceeding 100 frames per second, and a mean depth reconstruction error of 0.16 mm. The implementation provides a foundation for AI-assisted machining applications while preserving the machine tool monitoring and teleoperation capabilities.
\end{abstract}

\section{INTRODUCTION}
\label{sec:introduction}
A fundamental shift towards fossil-free steel production is ongoing to meet the long-term sustainability goals of the Paris Agreement~\cite{Paris2015agreement}. Currently, steelmaking accounts for about 7\% of global CO$_2$ emissions; to reduce the carbon footprint and support sustainable steelmaking, blast furnaces are replaced with electric arc furnaces, enabling the use of renewable energy sources~\cite{Pei2020toward}. In addition, ultra-strong steel materials are being developed to reduce the need for raw materials and the weight of end products.

\subsection{Motivation}
\label{subsec:motivation}
While innovative steel materials have the potential to significantly enhance the sustainability of end products, their adoption poses challenges for manufacturing companies. Advanced manufacturing and process optimization methods are required to efficiently fabricate sustainable products using ultra-strong, fossil-free steels~\cite{Jin2017impact}. Digital manufacturing tools supported by DTs and Artificial Intelligence (AI) provide opportunities to improve machining quality, productivity, and resource efficiency~\cite{Ntemi2022infrastructure}, paving the way for data-driven optimization of machining processes and facilitating the adoption of fossil-free and ultra-strong steels~\cite{Ward2021machining, Liu2023review}.

\subsection{Machine Tool Digital Twins}
\label{sec:machinestateoftheart}
Recent advances in DTs enable monitoring, simulation, and optimization of CNC machining processes~\cite{Ntemi2022infrastructure}. Existing approaches typically focus on tool monitoring, process simulation, or closed-loop machining process control. Although real-time machine synchronization has become available in commercial systems~\cite{Vericut2026}, integration of synchronized machine-level and machining process into a hierarchical DT remains limited.

Kaarlela and Outeiro introduced a CPMT framework to enable CNC teleoperation for workforce training~\cite{Kaarlela2025cyber}. The proposed framework established bidirectional communication between the physical machine tool and its digital counterpart, enabling real-time machine monitoring and control. Studer et al.~\cite{Studer2024openended} and Chuang et al.~\cite{chuang2024multiuser} implemented virtual training environments for milling machine operators. Both systems provide virtual machining environments without synchronization to a physical CNC machine, with Studer et al. focusing on single-user training and Chuang et al. extending the concept to collaborative multi-user training.

\subsection{Machining Process Digital Twins}
\label{sec:processstateoftheart}
Machining process DTs extend the machine tool DTs by representing the interaction between the cutting tools and the workpiece during material removal~\cite{Liu2023review}. High-fidelity process representations, including voxel- and dexel-based~\cite{Nie2024efficient, Dambly2022tridexel} workpiece models, have been proposed to estimate the evolving workpiece geometry and support adaptive machining and virtual verification. Ward et al.~\cite{Ward2021machining} proposed a machining DT based on real-time model-based simulations for closed-loop machining control. Their framework integrated machining simulations and CNC feedback to predict machining errors and optimize machining parameters in real-time. Dvorak et al. proposed a machining process DT for hybrid additive/substractive manufacturing~\cite{Dvorak2022machining}. The presented framework combines measurement data, milling stability, structural dynamics, cutting force, and geometrical models to support machining parameter selection, process verification, and quality assessment of the manufactured part. Cabral et al. established a DT framework for a legacy CNC mill using MQTT and serial communications. Their framework captures real-time controller and axis status and spindle loads to drive the DT visualizations~\cite{Cabral2023digital}.

\subsection{Research Gap}
\label{subsec:researchgap}
Existing machine tool DTs primarily focus on representing the operational state of CNC machines (machine tool geometry and kinematics) for monitoring, teleoperation, and operator training, whereas machining process DTs emphasize process simulation for toolpath verification and optimization. 

\begin{table}[!ht]
\caption{Comparison of machine tool and machining process DT frameworks.}
\label{tab:review}
\centering
\begin{tabular}{lccc}
\textbf{Publication} & \textbf{Year} & \textbf{Machine DT} & \textbf{Process DT} \\
\hline
Ward et al.~\cite{Ward2021machining}                 & 2021 & -- & \checkmark \\
Dvorak et al.~\cite{Dvorak2022machining}             & 2022 & -- & \checkmark \\
Cabral et al.~\cite{Cabral2023digital}               & 2023 & \checkmark & -- \\
Studer et al.~\cite{Studer2024openended}             & 2024 & -- & -- \\
Chuang et al.~\cite{chuang2024multiuser}             & 2024 & -- & -- \\
Kaarlela and Outeiro~\cite{Kaarlela2025cyber}        & 2025 & \checkmark & -- \\
\textbf{This work}                                  & 2027 & \checkmark & \checkmark \\
\hline
\end{tabular}
\end{table}
Existing machine tool and machining process DTs have been developed independently, rather than being integrated into a unified cyber-physical machining framework. This limitation motivates the development of a CPMT framework that simultaneously maintains synchronized CNC machine and machining process DTs together with a persistent Part DT Repository.

\subsection{Scientific Contribution}
\label{subsec:scientificcontribution}
This work extends the previously presented CPMT framework~\cite{Kaarlela2025cyber} into a hierarchical DT architecture that simultaneously maintains synchronized DTs of both the machine tool and the machining process. 

While preserving the machine tool DT for monitoring and teleoperation, the proposed framework introduces real-time synchronization of the evolving workpiece geometry, machining operation state, and vibration measurements. Furthermore, persistent storage of synchronized machining data enables process replay, workpiece traceability, and the future generation of synthetic datasets for AI model development.

\section{PROPOSED PROCESS DIGITAL TWIN CONCEPT}
\label{sec:proposedconcept}
The proposed framework consists of both a physical CNC machine and workpiece, their corresponding DTs, a communication and synchronization layer, and a Part DT Repository. Figure 1 illustrates the proposed CPMT framework. The full architecture is described in section~\ref{subsec:architecture}, The physical twin is described in Section~\ref{subsec:physicaltwin}, The architecture for the CNC and process DTs are described in Section~\ref{subsec:cncdigitaltwin} and \ref{subsec:processdigitaltwin} respectively. Finally, the communication layer is described in Section \ref{subsec:communications}.

\subsection{Architecture}
\label{subsec:architecture}
The presented approach is based on a cloud server architecture to enable location-independent access to the CNC machine control and monitoring. The cloud server provides the user with access to the DT runtime. The runtime is downloaded and executed by the user's web browser, and enables both desktop and immersive modes. The DT user interface is implemented as a Unity WebGL application hosted on the cloud server. Users access the system through a web browser without installing dedicated client software. 

\begin{figure*}[!tb]
  \centering
  \includegraphics[width=2\columnwidth]{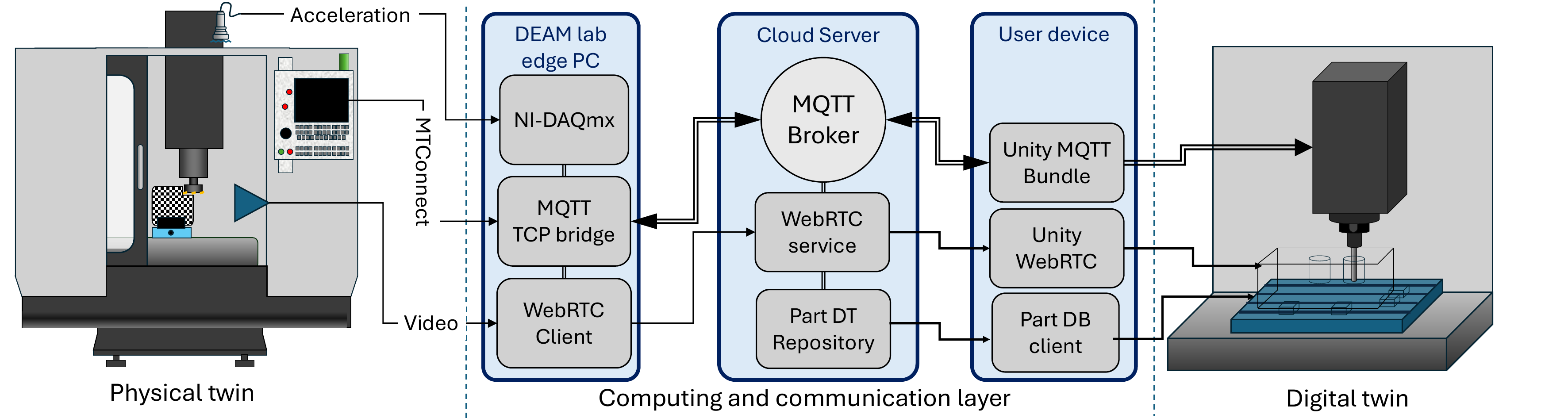}
  \caption{System architecture representing the physical twin, including CNC with Machine Tool Connect (MTConnect)~\cite{MTConnect2024} interface, acceleration sensor, and web camera. The computing and communication layer includes the edge PC, cloud server, the user device, and the virtual environment. The user device executes the Unity WebGL DT client received from the cloud server and communicates with cloud server backend services through Message Queuing Telemetry Transport (MQTT), Hypertext Transfer Protocol Secure (HTTPS), and Web Real-Time Communication (WebRTC).}
  \label{fig:architecture}
\end{figure*}

The Unity WebGL application supports both desktop and immersive modes through WebXR-compatible devices, enabling location- and device-independent monitoring and control of the CPMT. The cloud server also provides the communication layer broker and part DT repository services. The architecture is illustrated in Figure~\ref{fig:architecture}.  

The WebRTC service enables live video from the physical CNC machine to be displayed within the Unity WebGL DT environment as a texture overlaid on the workpiece surface. 

\subsection{Physical Twin} 
\label{subsec:physicaltwin}
The CNC physical twin is a 3-axis Haas Mini Mill CNC milling machine (Haas Automation, Inc., CA, USA)~\cite{HaasMinimill2024} located at the University of North Carolina at Charlotte. Detailed specifications are presented in Table~\ref{tab:haasspecs}. The milling tools utilized in this work are: (1) a 12 mm diameter ball-end mill, (2) a 10 mm diameter flat-end mill. 

\begin{table}[h!tb]
    \normalsize
    \centering
    \caption{Specifications of the milling machine.}
    \begin{tabular}{lc}
        \textbf{Item} & \textbf{Value}\\ \hline
        X-axis travel &  406 mm (16")\\ 
        Y-axis Travel &  356 mm (14")\\ 
        Z-axis Travel &  381 mm (15")\\ 
        Spindle Speed & 0-6000 rpm\\
        Linear velocity & 10 m/min (400 ipm)\\
        Positioning accuracy & 0,01 mm (0.0004")\\\hline 
    \end{tabular}
    \label{tab:haasspecs}
\end{table}

\subsection{CNC Machine Digital Twin} 
\label{subsec:cncdigitaltwin}
The CNC machine DT follows the same pipeline as in the previous work done by Kaarlela and Outeiro \cite{Kaarlela2025cyber}. This DT includes a representation of the physical milling machine, which has the geometry of the active tool, namely the tool envelope, which is the core tracked component used for material removal registration. The current implementation uses the two cutting tools mentioned above: (1) the ball-end mill, which sweeps a hemispherical tip along the toolpath to produce a capsule-shaped swept volume, and (2) the flat end mill, represented as a swept cylinder with a flat end. 

\subsection{Machining Process Digital Twin} 
\label{subsec:processdigitaltwin}
The Machining Process DT maintains a persistent digital representation of the workpiece throughout the machining operation. The current work introduces a voxel-based, periodically updated real-time representation of the workpiece. This includes a synchronized machine tool twin that is required to visualize the subtractive machining process, where each engagement of the tool removes material from the workpiece. Therefore, the digital representation of the workpiece needs to be updated as the material is removed. We adopt a volumetric representation that is well established in machining simulation~\cite{altintas2014virtual, joy2017frame}. The novelty of this work is not the representation itself but rather its use as a persistent workpiece-state estimate of the process twin driven by real-time machine data.

The full voxel-based representation is referred to as a stock, while a unit of the stock is referred to as a chunk, and each chunk consists of voxels. The stock is represented as a uniform voxel field over its axis-aligned bounding box. Each voxel stores a continuous scalar in [0,1] rather than a binary bit. Because this value varies smoothly across the surface, the marching cubes isosurface-extraction algorithm~\cite{bourke1994polygonising} interpolates the exact crossing point along each grid edge, placing the surface between voxel centers and recovering sub-voxel accuracy from a coarse grid. The surface mesh is then extracted using the marching cubes algorithm~\cite{bourke1994polygonising}, and vertex normals are taken from the gradient of the density field rather than recomputed from the triangle geometry, which follows the standard signed-distance formulation for implicit surfaces~\cite{osher2003level}. Storing a continuous scalar distinguishes the representation from a binary occupancy grid.

 Material removal is registered as a signed distance to the swept tool envelope. Each spawned voxel retains the minimum signed distance it has reached over the course of machining, and its density is derived from this single accumulated distance field. Retaining one authoritative distance per voxel reduces visible seams and yields a consistent surface regardless of the order or overlap of the cuts that produced it.

\begin{figure}[htbp]
  \centering
  \includegraphics[width=\linewidth, trim = {8cm 0 0 0}, clip]{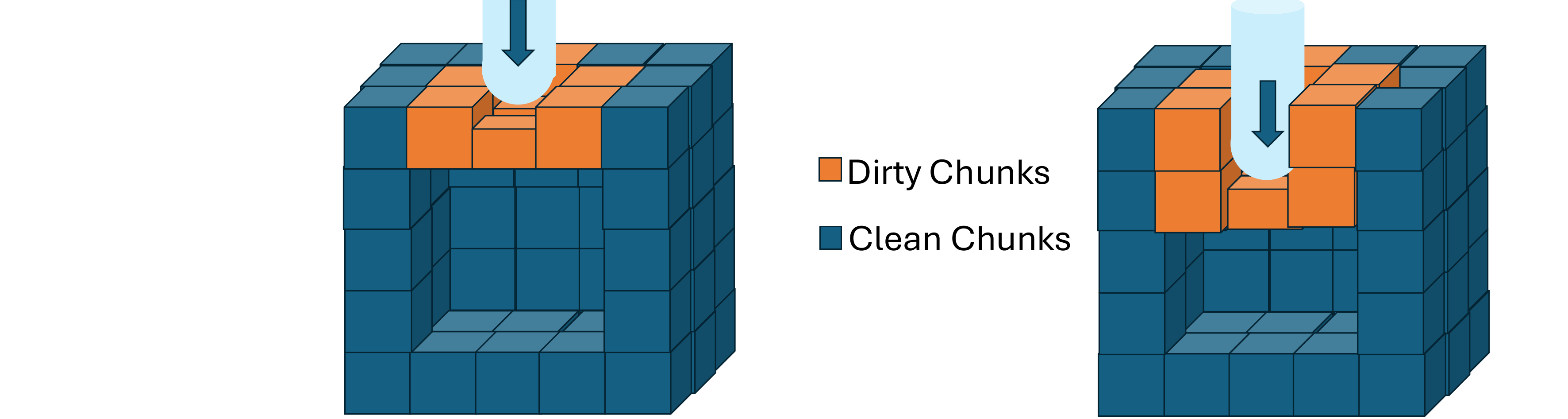}

  \begin{subfigure}[t]{0.48\linewidth}
    \centering
    \caption{Initial stock state}
    \label{fig:Intial-stock-state}
  \end{subfigure}
  \hfill
  \begin{subfigure}[t]{0.48\linewidth}
    \centering
    \caption{Lazy generation as mill goes deeper.}
    \label{fig:Lazy-Generation}
  \end{subfigure}

  \caption{Stock state as mill cuts through; dirty chunks are regenerated every frame while clean chunks are ignored.}
  \label{fig:drill}
\end{figure}

The voxel-system representation is adopted over the dexel-based schemes, which represent the stock as a field of depth-elements sampled along a fixed direction, despite its higher memory costs~\cite{joy2017frame, boz2015comparison}, due to its general purpose and update robustness and accuracy against rotational cuts~\cite{boz2015comparison, schnos2021gpu}. To maximize stock resolution while minimizing computational cost, we initially render only the outer shell of the stock (Fig.~\ref{fig:Intial-stock-state}) and instantiate the inner chunks only when the end mill reaches them (Fig.~\ref{fig:Lazy-Generation}). Further batching and a bounded per-frame rebuild budget are available. This lazy chunk generation strategy follows established practices in graphics and visual effects~\cite{schnos2021gpu}. 

\subsection{Communication and Synchronization} 
\label{subsec:communications}
The optional software feature Machine Data Collection (MDC)~\cite{HaasMDC2024} enables MTConnect to read machine data and a low-level Q and E TCP command interface for writing machine tool data. The CNC connects to the cloud-based CPMT computation using these interfaces to enable monitoring and teleoperation of the CNC.  

The communication layer is based on the MQTT publisher/subscriber protocol. On the DT, the Unity MQTT Bundle add-on~\cite{MQTTbundle2024} enables the communication. A custom Python TCP-to-MQTT application was written for the edge PC to generate HTTPS requests to collect, parse, and publish the operational data from the CNC and to time-synchronize the collected operational and vibration data. The vibration data is collected in the background, and a timestamped MQTT data package is generated each time MTConnect operational data is received. Listing~\ref{lst:sample_python} details the JavaScript Object Notation (JSON) message format.

\begin{lstlisting}[language=json,numbers=none,label={lst:sample_python}, caption={Process DT message.}, basicstyle=\scriptsize\ttfamily]
{"type": "REPLAY_STATE",
  "timestamp_epoch": 1778839083.917025,
  "toolno": 1,
  "x": -164.887,
  "y": -130.986,
  "z": -155,
  "spindle_rpm": 1997.46,
  "tool_length": 50,
  "tool_radius": 16,
  "offset_x": -200,
  "offset_y": -200,
  "offset_z": -250,
  "vibration": {
    "dt_s": 0.278213,
    "f_dom_hz": 34.91,
    "amp_rms_g": 0.2403,
    "amp_peak_g": 0.3399,
    "overall_rms_g": 1.0215,
    "overall_peak_g": 9.5309}}
\end{lstlisting}

The forward path (machine $\rightarrow$ DT), in which machine state is published over MQTT and consumed by the Unity DT, utilizes MTConnect. Since MTConnect provides read-only access \cite{Kaarlela2025cyber} to the machine data, the return path (DT $\rightarrow$ machine) is implemented as a separate command channel utilizing TCP commands. The Unity client publishes JSON command messages to the MQTT broker, and a Python command bridge subscribes to them and translates them into a machine action through the CNC command interface. By choice, all command parsing and command translation is done on the edge PC. This design choice allows Unity to handle only flat JSON messages in both directions, which preserves MTConnect as the semantic model for monitoring while still enabling the bidirectional control loop.

\subsection{Part Digital Twin Repository}
\label{subsec:manufacturingdigitaltwinrepository}
The proposed framework introduces a Part DT Repository to persistently store manufactured workpiece information. The repository is implemented using MariaDB hosted on the cloud server, enabling centralized management of part-specific DT information and remote access for authorized users.

For each machining operation, the repository stores synchronized CNC operational data, workpiece geometry, active tool information, machining parameters, vibration measurements, and timestamps. The stored information represents the complete manufacturing history of an individual workpiece and is associated with its corresponding part identifier.

\subsection{Cybersecurity}
\label{subsec:cybersecurity}
The DEAM Lab edge PC connects to the Internet using a firewall configured to allow access only on port 8883 for secure, encrypted, and authenticated MQTT communications. Disabling access to all other ports prevents external users from executing low-level TCP commands on the CNC. 

The cloud server is protected by a software firewall that permits access only to the ports necessary for MQTT, HTTPS, and the Secure Shell (SSH) services. Data exchanged over MQTT and HTTPS is secured using a CA certificate issued by GEANT Vereniging. Both the Apache web server and the Mosquitto MQTT broker are configured to allow exclusively encrypted client connections.

To maintain the security of the deployed framework over time, automated vulnerability scans are periodically performed on both the cloud server and the DEAM laboratory edge PC. These scans identify outdated firmware and software components and notify system administrators of potential vulnerabilities. When vulnerabilities are detected, mitigation measures include upgrading the affected software to a secure version, reverting to a previous secure version when appropriate, or disabling the affected service until a secure update becomes available.

\section{RESULTS AND DISCUSSION}
\label{sec:results}
The main result of this work is a hierarchical DT framework that maintains synchronized digital representations of both the CNC machine and the evolving workpiece during machining. The framework integrates machine states, workpiece geometry, and synchronized vibration measurements into a unified machining process representation. Figure~\ref{fig:workflow} illustrates the machining process DT.

\subsection{Machining Process Digital Twin}
\label{subsec:processdt}
\begin{figure}[h!tb]
    \centering
    \begin{subfigure}[t]{0.48\columnwidth}
        \centering
        \includegraphics[width=\linewidth]{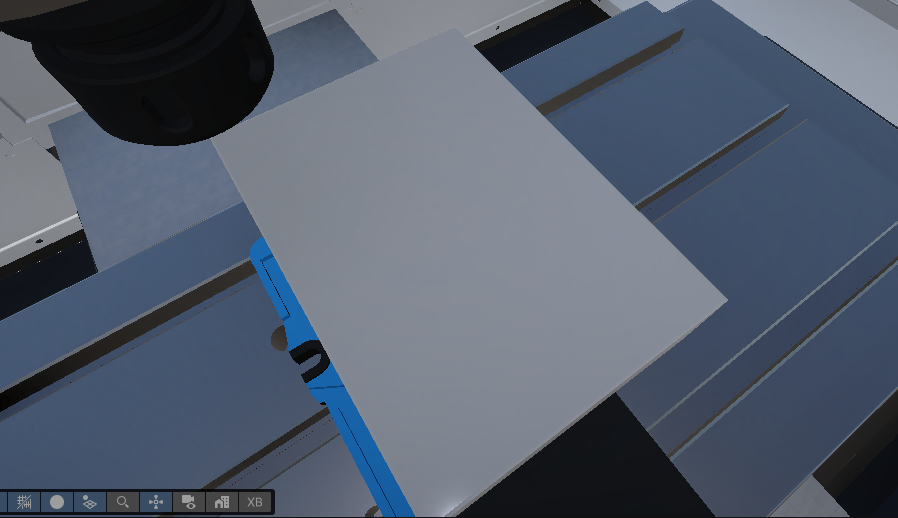}
        \caption{Prior to machining.}
        \label{fig:firstW}
    \end{subfigure}
    \hfill
    \begin{subfigure}[t]{0.48\columnwidth}
        \centering
        \includegraphics[width=\linewidth]{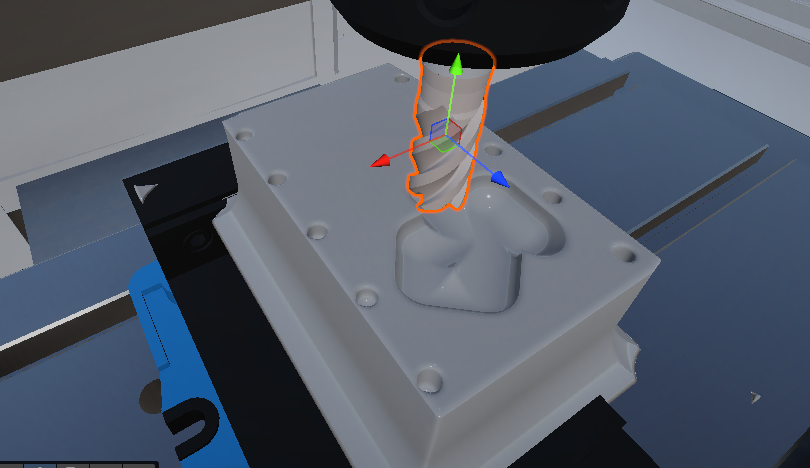}
        \caption{During machining cycle.}
        \label{fig:secondW}
    \end{subfigure}
    \vspace{0.5em}
    \begin{subfigure}[t]{0.48\columnwidth}
        \centering
        \includegraphics[width=\linewidth,  trim = {0 0 0 0}, clip]{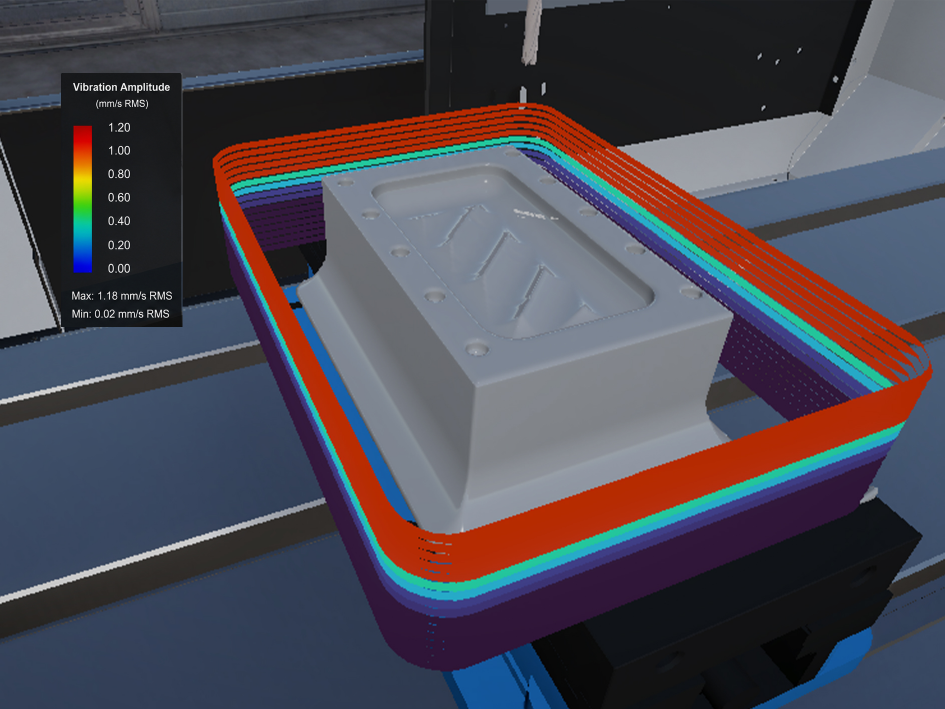}
        \caption{Vibration overlay.}
        \label{fig:thirdW}
    \end{subfigure}
    \hfill
    \begin{subfigure}[t]{0.48\columnwidth}
        \centering
        \includegraphics[width=\linewidth,  trim = {0 0 0 1.6cm}, clip]{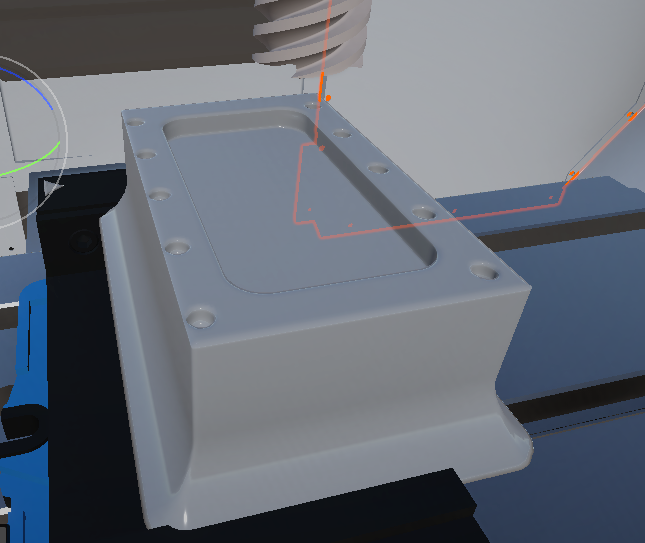}
        \caption{Machining finished.}
        \label{fig:fourthW}
    \end{subfigure}

        \caption{Illustration of the process DT in different phases of the machining. }
    \label{fig:workflow}
\end{figure}

The implemented framework maintained synchronized digital representations of both the CNC machine and the evolving workpiece throughout the machining cycle. Figure~\ref{fig:workflow} illustrates the process DT before machining, during material removal, with synchronized vibration information, and after completion of the operation. The voxel-based digital stock model was updated using the measured tool position and active tool geometry received from the CNC, enabling the workpiece twin to evolve during the physical machining process.

The process DT starts with a voxel volume of 125 $\times$ 175 $\times$ 50~mm with a baseline grid of 160 $\times$ 224 $\times$ 64 voxels of 0.78125~mm, a factor of four in linear resolution and 64 in voxel count, and a machining-state update rate of 20~Hz. On a computer equipped with Intel(R) Core(TM) Ultra 7 265H,  Intel(R) Arc(TM) 140T GPU, and 64~GB of RAM, the WebGL runtime maintained  an average frame rate of 110.4~frames/s during machining, with a median frame time of 8~ms and a 1\% low frame rate of
49.1~frames/s. Lazy chunk generation limited processing to the regions affected by the cutting tool. 

\begin{table}[!h]
\caption{Interactive performance across configurations. Values are means over $n$ captures.}
\label{tab:perf}
\centering
\footnotesize
\setlength{\tabcolsep}{4pt}
\begin{tabular}{ccccccc}
\textbf{Configuration} & \textbf{Chunks} & \textbf{$n$} & \textbf{FPS} & \textbf{1\% low} & \textbf{$p_{99}$~(ms)} & \textbf{Hitches}\\
\midrule
\multicolumn{7}{l}{\emph{Chunk size, baseline grid $160\times224\times64$}}\\
\quad 4  & 35\,840 & 2 & 39.0  & 10.7 & 80 & 811\\
\quad 16 & 560     & 3 & \textbf{110.4} & \textbf{49.1} & \textbf{18} & \textbf{6}\\
\quad 32 & 70      & 3 & 81.4  & 19.9 & 43 & 175\\
\midrule
\multicolumn{7}{l}{\emph{Voxel resolution, chunk size 4}}\\
\quad Coarse   & 560     & 1 & 120.0 & 87.3 & 10  & 0\\
\quad Low      & 4\,480  & 3 & 43.0  & 22.1 & 41  & 617\\
\quad Baseline & 35\,840 & 2 & 39.0  & 10.7 & 80  & 811\\
\quad High     & 286\,720& 1 & 2.3   & 1.7  & 559 & 95\\
\bottomrule
\end{tabular}
\end{table}
 
Table~\ref{tab:perf} summarizes recorded performance across different configurations. Chunk granularity, rather than voxel resolution, was found to be the dominant tuning parameter. At the baseline grid, increasing the chunk size from 4 to 16 voxels per chunk per axis increased the average frame rate from 39.0 to 110.4~frames/s and reduced the recorded frame hitches from 811 to 6, while a further increase to 32 voxels per chunk per axis reduced performance to 81.4~frames/s mainly due to the number of voxels needed to be rebuilt at 32 voxels per chunk. Large chunks invert the trade-off: the median frame time at chunk sizes 16 and 32 is identical at 9~ms, but the 99th-percentile frame time differs by a factor of 2.4, because a single cut dirties a chunk of 32\,768 voxels rather than 4\,096. 

The distribution is more informative than the average throughout. The chunk sizes of 16 and 32 are indistinguishable by median frame time yet differ by a factor of 29 in hitch count, since the cost of surface reconstruction is concentrated in the frames in which the tool crosses a chunk boundary rather than distributed across the capture.

Voxel resolution sets the second bound. Across the resolution sweep, the coarse configuration sustained 120.0~frames/s without recorded hitches, whereas the high configuration collapsed to 2.3~frames/s with a median frame time of 422~ms under cutting, having rendered at 36.7~frames/s while idle; the cost is therefore attributable to surface reconstruction rather than to rendering.

The geometric fidelity of the reconstructed workpiece was evaluated by probing the density field at the completed hole locations produced by a single tool and comparing the reconstructed depth against the commanded depth. Across ten drilled features with a nominal depth of 50.00~mm, the mean absolute depth error was 0.16~mm, equivalent to 0.20 voxels, with a maximum error of 0,67~mm. Nine of the ten features fell within the 0.39~mm bound imposed by the $\pm\frac{1}{2}$ voxel discretization limit at the baseline resolution. 

The synchronized data package combined the CNC axis positions, spindle speed, active tool information, and vibration measurements using a common timestamp. The resulting process history was stored in the Part DT Repository and could be retrieved to replay the machining operation.

The results demonstrate that the proposed framework can maintain a persistent estimate of the evolving workpiece while preserving the monitoring and teleoperation capabilities of the CNC DT. The lazy chunk generation strategy reduced unnecessary mesh reconstruction by updating only the regions affected by material removal. However, the achievable geometric fidelity remains dependent on the selected voxel resolution and the update frequency of the CNC data stream.

The persistent storage of synchronized machining and geometry data enables replay of completed machining operations, traceability of individual workpieces, and generation of synthetic DT datasets by replaying, perturbing, and simulating machining scenarios for AI training and validation.

\subsection{Current Limitations}
\label{subsec:limitations}
One limitation follows from this representation. A single scalar per voxel admits at most one tool crossing per grid edge, which means that cuts thinner than a voxel may be missed. This is the case addressed by the two-parameter frame-crossing points \cite{joy2017frame}, which recovers such features at additional memory costs. At the current scale, such features do not arise, and the single-scalar field is retained for its simplicity and lower memory footprint.

The reconstruction is therefore accurate to the discretization floor of the representation, and the dominant error term is the voxel size rather than the synchronization pipeline.

The synchronization fidelity of the framework is limited by the MTConnect data acquisition approach. Since machine states are obtained through periodic HTTP requests rather than continuous streaming, intermediate tool positions are not captured by the DT. The active frequency for the HTTP requests is set to 20~Hz, which was selected to eliminate incorrect material removal due to missed tool poses at the previous rate of 2~Hz. Increasing the update rate introduces the limit of the client having to reconstruct the affected region of voxels within the interval between samples. This increase in update rate has to be balanced against transport capacity, which risks drifting end-to-end latency with operating time. Consequently, sampling remains discrete, so trajectory detail between samples is not represented; this bounds the fidelity achievable for operations whose duration is short relative to the sampling interval. 


Another practical constraint is that the simulation program specifies tool positions and the origin of the workpiece, but never specifies the dimensions of the workpiece. This introduces an automated calibration challenge, which is currently bypassed by a manual adjustment of the workpiece dimensions.

\subsection{Future Work}
\label{subsec:futurework}
Future work will integrate the NC machining program and AI-based trajectory state estimation into the DT framework. By combining the programmed toolpath with the measured machine states, the CPMT framework can reconstruct and estimate intermediate tool positions that are not directly captured through MTConnect. This will improve the continuity of the machining process representation and enable a more precise synchronization of process measurements, workpiece evolution, and machine-tool states. WebRTC-based projection of captured surface imagery onto the Part DT will be studied to visualize surface quality and roughness.

Beyond trajectory state estimation, AI has the potential to detect chatter, optimize the machining parameters, and estimate the tool wear. The Part DT Repository enables generation of synthetic machining datasets that can be used to train and validate AI models for process monitoring, anomaly detection, quality prediction, tool wear estimation, and adaptive process optimization. 

\section{CONCLUSIONS}
\label{sec:conclusions}
This paper presented a CPMT framework that extends a previously developed CNC DT with a synchronized Machining Process DT. The proposed framework integrates real-time CNC operational data, a voxel-based representation of the evolving workpiece, synchronized vibration measurements, and a cloud-based Part DT Repository. Together, these components provide a unified digital representation of both the machine tool and the machining process throughout the manufacturing operation.

The framework enables synchronized visualization and persistent management of machining process information for replay and traceability.

Future work will focus on validating the framework in industrial machining scenarios, integrating additional process sensing modalities such as cutting forces, surface quality, and chip formation, and investigating AI-driven process optimization using recorded and synthetically generated machining datasets derived from the Part DT Repository. These capabilities are expected to support efficient machining of ultra-strong and fossil-free steels through data-driven optimization of machining processes.






\section*{ACKNOWLEDGMENT}

This research was supported by RIS4E - Revolutionary and 
Intelligent Steel Solutions for Sustainable Environment 
(Business Finland 43/31/2026), and by the University of North Carolina at Charlotte through the Center for Precision Metrology Affiliates Program.

\bibliographystyle{IEEEtran}
\bibliography{citations}

\end{document}